\documentclass{PoS}

\usepackage{graphicx}
\usepackage{amsmath}
\usepackage{url}
\usepackage{color}
\usepackage{cite}
\usepackage{subfig}
\usepackage{bm}
\usepackage{float}

\newcommand{\GeVc}{GeV/$c$}
\newcommand{\GeVcc}{GeV/$c^2$}

\newcommand{\beq}{\begin{equation}}
\newcommand{\eeq}{\end{equation}}
\newcommand{\dif}[1]{\mathrm{d} #1 \,}

\newcommand{\mr}[1]{\mathrm{#1}}
\newcommand{\refeq}[1]{Eq.~(\ref{#1})}

\newcommand{\phiS}{\varphi_\mr{S}}
\newcommand{\kT}{$k_\mr{T}$}

\newcommand{\PhTscal}{P_\mr{T}}
\newcommand{\qT}{\vec{q_\mr{T}}}
\newcommand{\qTscal}{q_\mr{T}}
\newcommand{\qTscalt}{$q_\mathrm{T}$}
\newcommand{\kaT}{\vec{k_{a\mr{T}}}}

\newcommand{\kbT}{\vec{k_{b\mr{T}}}}

\newcommand{\fSiv}[1]{f_{1\mr{T}}^{\perp #1}}
\newcommand{\fSivb}[1]{f_{1\mr{T},b}^{\perp #1}}
\newcommand{\fSivp}[1]{f_{1\mr{T},\mr{p}}^{\perp #1}}
\newcommand{\F}[2]{F_\mr{#1}^{#2}}
\newcommand{\A}[2]{A_\mr{#1}^{#2}}
\newcommand{\qqswap}{(q\leftrightarrow\bar{q})}

\title{Measurement of $\mathbf{q_T}$-weighted transverse-spin-dependent azimuthal asymmetries at COMPASS}

\ShortTitle{Measurement of $q_T$-weighted TSAs at COMPASS}

\author{\speaker{Riccardo Longo}  \thanks{On Behalf of the COMPASS collaboration}\\
        University of Illinois at Urbana-Champaign \\
        E-mail: \email{ riccardo.longo@cern.ch }}

\abstract{COMPASS is a fixed-target experiment in operation at the CERN North Area (SPS, M2 beam-line) since 2002. An important part of the broad physics programme of the experiment  is dedicated to the exploration of the transverse spin-structure of the nucleon studying target transverse spin dependent azimuthal asymmetries (TSAs) arising in the Semi-Inclusive DIS (SIDIS) and Drell-Yan (DY) cross-sections. In addition to those measurements, COMPASS has recently studied also the TSAs weighted by powers of the hadron transverse momentum (in SIDIS) and virtual photon transverse momentum, $q_T$ (in DY). In the transverse momentum dependent (TMD) QCD approach, the conventional DY TSAs are interpreted as convolutions of the beam pion and of the transversely polarized target proton TMD parton distribution functions (PDFs), while the \qTscalt-weighted TSAs can be interpreted as simple products of transverse moments of the TMD PDFs.

In 2015 and 2018 COMPASS performed two years of Drell-Yan data taking with a 190 GeV/$c$ $\pi^-$ beam impinging on a transversely polarized NH$_3$ target. The analysis of the \qTscalt\-weighted TSAs performed on these two data sets is presented in this paper. The results for DY Sivers $q_T$ weighted TSA are compared with the expectations based on the studies of the weighted Sivers asymmetry measured in the SIDIS process. Combining the information from SIDIS and DY measurements, the pion Boer-Mulders TMD PDF is also studied. }

\FullConference{XXVII International Workshop on Deep-Inelastic Scattering and Related Subjects - DIS2019\\
		8-12 April, 2019\\
		Torino, Italy}

\begin{document}

\section{Introduction}
The structure of a polarized nucleon, within the "twist-2" approximation of the QCD parton model, is described by a set of eight transverse-momentum-dependent (TMD) parton distribution functions (PDFs). They encode all possible correlations between the nucleon spin, the parton spin and the transverse component of the intrinsic parton momentum, \kT. Each TMD PDF depends on the fraction $x$ of the nucleon momentum carried by the parton and its \kT. 
A powerful method used to access these TMD PDFs is the study of transverse spin-dependent azimuthal asymmetries (TSAs) arising in Semi-Inclusive Deep Inelastic Scattering (SIDIS) and Drell-Yan (DY) cross-sections. 
The TMD-factorization theorem has been proven for SIDIS and DY cross-sections \cite{Collins:2011zzd}, allowing for the interpretation of the TSAs as convolutions of TMD PDFs and TMD fragmentation functions in the case of SIDIS, or nucleon and incoming hadron TMD PDFs in the case of DY.
The idea of weighted TSAs was first brought up in the context of SIDIS\,\cite{kotzinian:1996,boer:1998}, and then ported to DY as well\,\cite{sissakian:2005b,wang:2017}. 
In both SIDIS and DY processes, in order to extract the TMD PDFs they have to be decomposed from the convolution integrals.
In the case of weighted TSAs, those convolutions are replaced by products of transverse moments of the TMD PDFs. 
 
Last year, COMPASS complemented a series of measurements of the SIDIS TSAs by publishing the weighted Sivers asymmetries extracted from the same data sample\,\cite{compass:2018weighted}. An identical approach was followed for the 2015 Drell-Yan analysis where, in addition to the standard TSAs\,\cite{compass:2017dy}, the corresponding virtual photon transverse momentum, $q_T$, weighted  TSAs\,\cite{matousek:2017} were also extracted.  The DY results have been updated after the analysis of $\sim$50$\%$ of the data collected in 2018, discussed in this paper. 

\begin{figure}[t]
	\begin{center}
	\subfloat[]{
		\label{fig:TRF}
		\includegraphics[width=0.33\textwidth]{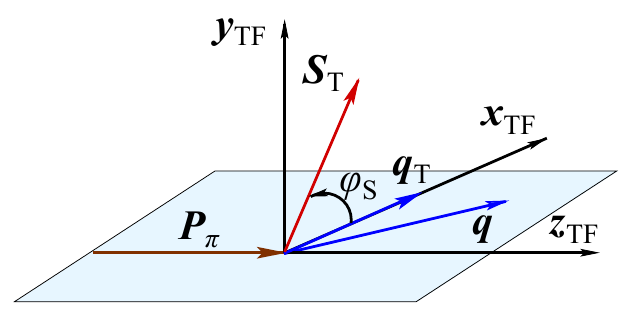}
		}\hspace{0.5cm}
	\subfloat[]{
		\label{fig:CS}
		\includegraphics[width=0.33\textwidth]{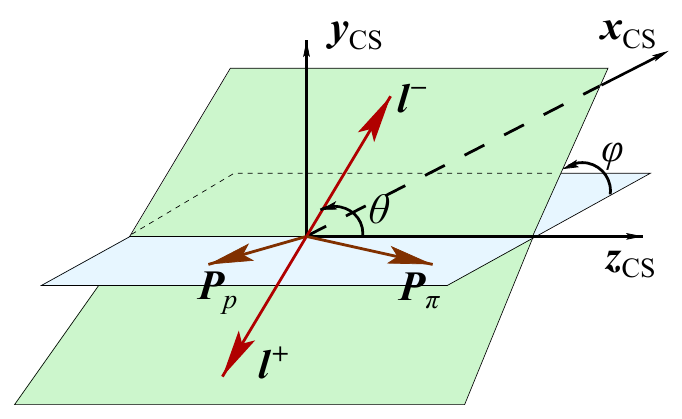}
		}
	\caption{The reference frames used: (a) the target frame (target particle rest frame), (b) Collins--Soper frame (dilepton rest frame), and the definition of the angles $\theta$, $\varphi$, and $\phiS$.}
	\label{fig:frames}
	\end{center}
\end{figure}

In both 2015 and 2018 data taking, COMPASS performed Drell-Yan measurements ($\pi^- \mr{p}^\uparrow \rightarrow \mu^- \mu^+ X$) induced by a 190  \GeVc\ $\pi^-$ beam scattering off a transversely polarized  NH$_3$ target. At leading order, the corresponding single-polarized Drell-Yan cross-section can be written as \cite{arnold:2008}:
\beq
	\label{eq:dy_xsec_lo}
	\begin{split}
		\frac{\dif{\sigma_\mr{DY}}}
		{\dif{x_\pi}\dif{x_N}\dif{\qTscal^2}\dif{\phiS}\dif{\cos\theta}\dif{\varphi}}
		\propto \biggl\{ &
			  (1+\cos^2\theta) \F{U}{1}
			+ \sin^2\theta \, \cos2\varphi \, \F{U}{\cos2\varphi} \\
		&	+ |\vec{S}_\mr{T}| \biggl[
				  (1+\cos^2\theta) \sin\phiS \, \F{T}{\sin\phiS} \\
		& \qquad + \sin^2\theta \, \sin(2\varphi+\phiS) \F{T}{\sin(2\varphi+\phiS)} \\
		& \qquad + \sin^2\theta \, \sin(2\varphi-\phiS) \F{T}{\sin(2\varphi-\phiS)}
				\biggr]
			\biggr\},
	\end{split}
\eeq
where $F_X^{\mr{[mod]}} = F_X^{\mr{[mod]}}(x_\pi,x_N,\qTscal)$ are the structure functions\footnote{In the notation $F_X^{\mr{[mod]}}$ the superscript $\mr{[mod]}$ indicates the associated modulation, while the subscript $X$ denotes the target polarization states ("U" stands for unpolarized and "T" for transverse polarization). }, $\varphi$ ($\theta$) represents the azimuthal (polar) angle of the lepton momentum in the Collins-Soper frame (Fig.~\ref{fig:CS}) and $\phiS$ the azimuthal angle of the target spin vector in the target rest frame (Fig.~\ref{fig:TRF}).

In Eq.~\ref{eq:dy_xsec_lo} one can identify five different terms, each containing an orthogonal modulation in $\phi$ or $\phiS$ and a structure function $\F{X}{\mr{[mod]}}$. The standard TSAs are defined as ratios between structure functions, $\A{U/T}{\mr{[mod]}} = \F{U/T}.{\mr{[mod]}} / \F{U}{1}$. The structure functions can be written as a flavour sum of convolutions of TMD PDFs over the intrinsic momenta of the two colliding quarks $\kaT$ and $\kbT$.  Therefore, measuring the TSAs gives access to a ratio of convolutions of TMD PDFs. The convolutions are usually solved assuming a certain \kT-dependence of the TMD PDFs (e.g. a Gaussian parameterization). 

The weighted TSAs, on the other hand, represent a way to avoid making any assumption on the \kT-dependence. They take advantage of the fact that, if one integrates the structure functions over $\qT =\vec{k}_{aT}$ and $\vec{k}_{bT}$ with appropriate weights $W_{X}$, the convolutions can be easily solved. 
The generic \qTscalt-weighted TSA can be written as 
\beq \A{T}{X W_X} = \frac { \int \dif{^2\qT} W_X \, \F{T}{X}}{ \int \dif{^2\qT} \F{U}{1} } \eeq
Considering the specific case of pion-induced Drell-Yan on a transversely polarized proton target, the three \qTscalt -weighted TSAs accessible are 
\begin{align}
    \label{eq:siv}
    \A{T}{\sin\phiS \frac{\qTscal}{M_\mr{p}}}(x_\pi,x_N)
			&= - 2 \frac{\sum_q e_q^2 \bigl[ f_{1,\pi^-}^{\bar{q}}(x_\pi) \, 
						\fSivp{(1)q}(x_N) + \qqswap \bigr]}
				 {\sum_q e_q^2 
					\bigl[ f_{1,\pi^-}^{\bar{q}}(x_\pi)\, 
						   f_{1,\mr{p}}^q(x_N) + \qqswap \bigr]}
    \\
    \label{eq:transv}
	\A{T}{\sin(2\varphi-\phiS) \frac{\qTscal}{M_\pi}}(x_\pi,x_N)
			&= - 2 \frac{ \sum_q e_q^2 \bigl[ 
							h_{1,\pi^-}^{\perp (1) \bar{q}}(x_\pi) \, 
							h_{1,\mr{p}}^q (x_N) 
						+ \qqswap \bigr]}
				 {\sum_q e_q^2 
					\bigl[ f_{1,\pi^-}^{\bar{q}}(x_\pi) f_{1,\mr{p}}^q(x_N) + \qqswap \bigr]}
    \\
    \label{eq:pretz}
	\A{T}{\sin(2\varphi+\phiS) \frac{\qTscal^3}{2M_\pi M_\mr{p}^2}}(x_\pi,x_N)
			&= - 2 \frac{ \sum_q e_q^2 \bigl[ 
							h_{1,\pi^-}^{\perp (1) \bar{q}}(x_\pi) \, 
							h_{1\mr{T,p}}^{\perp (2) q} (x_N) 
						+ \qqswap \bigr]}
				 {\sum_q e_q^2 
					\bigl[ f_{1,\pi^-}^{\bar{q}}(x_\pi) f_{1,\mr{p}}^q(x_N) + \qqswap \bigr]}
\end{align}
where the sums run over all quarks and antiquarks flavours $q$ with fractional electric charge $e_q$. $M_{\pi}$ and $M_{\mr{p}}$ represent the pion and proton masses and $f^{(n)}$ or $h^{(n)}$ are the $n$-th \kT$^2$-moments of the TMD PDFs.
The comparison of the weighted Sivers asymmetries measured in SIDIS and Drell-Yan processes has been proposed as an easy way to compare magnitudes and signs of the effect in the two processes~\cite{efremov:2004}. 

\section{Data collection and analysis}
\label{sec:measurement}
The analysis presented in this Letter is based on the Drell-Yan data collected by COMPASS in the year 2018, in essentially the same conditions of 2015. 
More details on the measurement can be found in \cite{bakur:2019dis}.
The 23 weeks of data-taking were divided in 9 periods, each consisting of consecutive weeks with opposite target polarizations. In this Letter, only $\sim$50$\%$ of the available 2018 data have been analyzed. 
Contrary to the standard TSAs analysis, no cut on \qTscalt\ is applied since the integration over the full \qTscalt\ range, where the cross-section is non-zero, is essential for the \qTscalt-weighted TSAs. Instead, a cut on the transverse momentum of the reconstructed muons, $\ell_{T}^{\mu_{\pm}} \, < \, 7$ \GeVc, was applied to remove badly reconstructed events.

\begin{figure}[!t]
	\begin{center}
	\subfloat[]{
		\label{fig:xbxN}
		\includegraphics[width=0.47\textwidth]{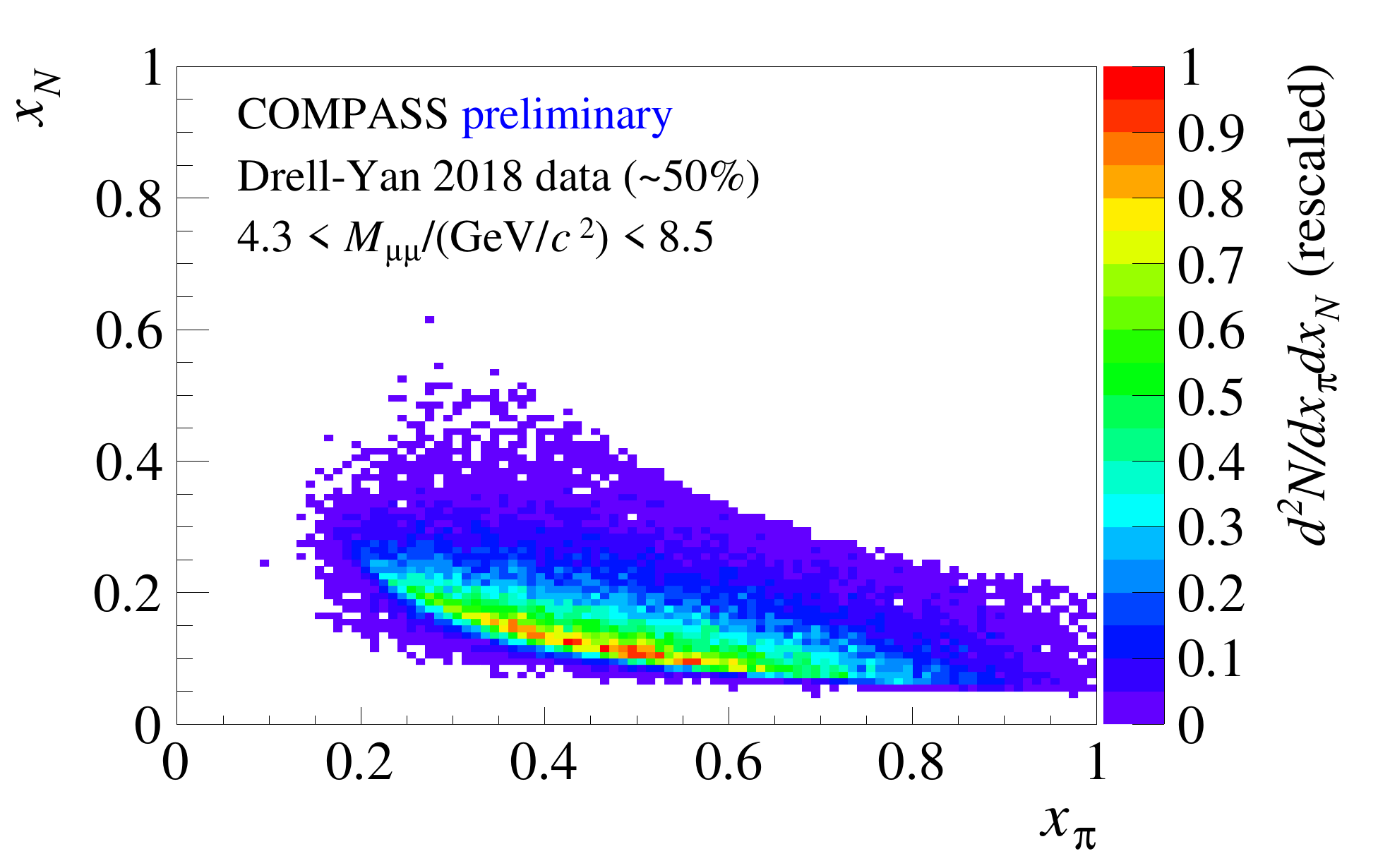}
		}\hspace{0.5cm}
	\subfloat[]{
		\label{fig:qt2018}
		\includegraphics[width=0.32\textwidth]{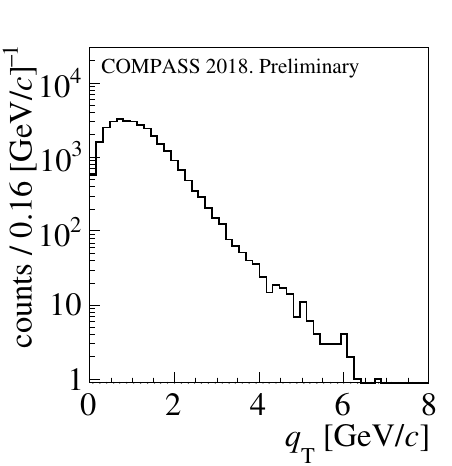}
		}
	\caption{$x_{N}$ versus $x_{\pi}$ (a) and $q_T$ (b) distributions  in the HM range from NH$_3$, 2018 data.}
	\label{fig:kine2018}
	\end{center}
\end{figure}

\begin{figure}[!h]
\centering
    \includegraphics[width=0.85\textwidth]{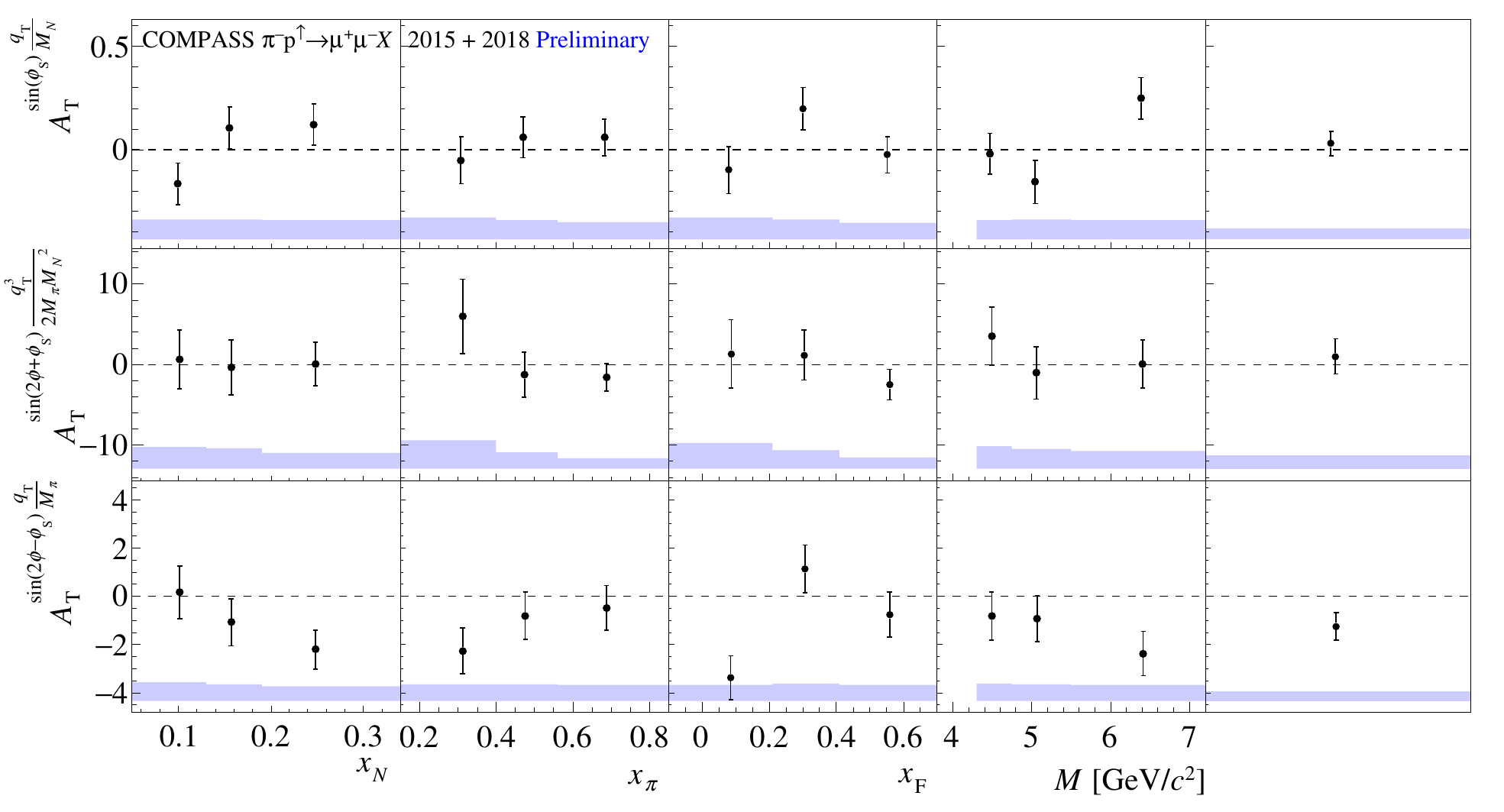}
    \caption{\label{fig:wasymcom} The \qTscalt-weighted TSAs obtained combining the whole 2015 data set with $\sim$ 50$\%$ of 2018 data set. The systematic uncertainties are represented by the blue bands. Normalization uncertainties from target polarization (5\,\%) and dilution factor calculation (8\,\%) are not shown.}
\end{figure}

As done for the standard TSAs, the \qTscalt-weighted TSAs are extracted in the so-called \textit{High-Mass} (HM) range, $M_{\mu\mu}  \in [ 4.3, 8.5 ] $ \GeVcc, which is particularly suited to study the predicted sign-change of the Sivers TMD PDF, as explained in \cite{compass:2017dy}. 
$x_N$ vs $x_\pi$ and \qTscalt\ distributions for the selected 2018 sample are shown in Fig.~\ref{fig:kine2018}. 
Each weighted TSA is obtained from fits to the data using the so-called 'modified double ratio method' \cite{matousek:2018}. 
The results obtained from the combined 2015 and 2018 samples are presented in Fig.~\ref{fig:wasymcom}. Systematic uncertainties are dominated by the impact of a possible imperfection of the acceptance cancellation, estimated evaluating the so-called `false asymmetries'. 

\section{Transverse momentum weighted asymmetries in SIDIS and Drell--Yan}
\subsection{Sivers asymmetry}
\begin{figure}
	\begin{center}
	\subfloat[]{
		\label{fig:ResSivPredDY}
		\includegraphics[width=0.4\textwidth]{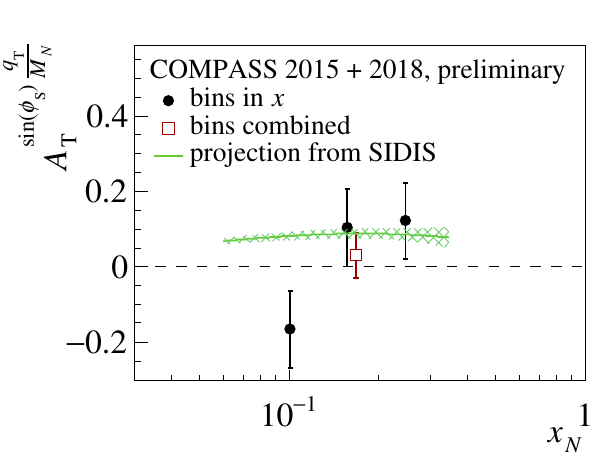}
		}
	\subfloat[]{
		\label{fig:ResBMDY}
		\includegraphics[width=0.4\textwidth]{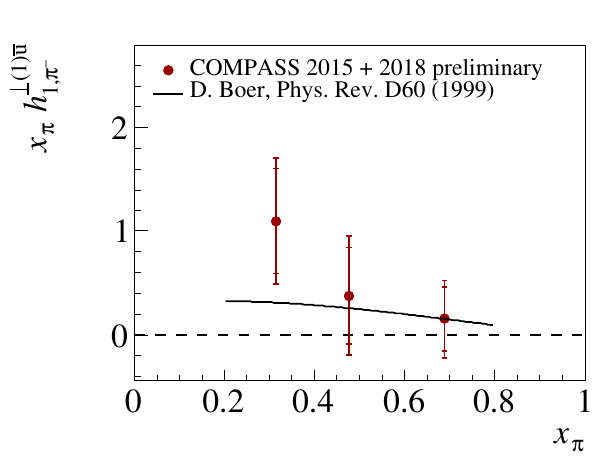}
		}
	\caption{(a) The weighted Sivers asymmetry extracted from COMPASS 2015 and 2018 ($\sim$ 50$\%$) DY data, compared with the 
    	projection from SIDIS (assuming sign-change). (b) The Boer-Mulders function of the pion as extracted from COMPASS 2015 and 2018 ($\sim$ 50$\%$)  data compared with the parametrization of Ref.~\cite{boer:1999 }.}
	\label{fig:results}
	\end{center}
\end{figure}

Compared to the standard TSAs, the weighted asymmetries represent a complementary and more straightforward way to compare the results obtained by COMPASS in SIDIS and in DY.  Recently, COMPASS has published the $\PhTscal/z$-weighted Sivers asymmetry in the production of charged hadrons in SIDIS \cite{compass:2018weighted}.
The first $k_T^2$-moment of the Sivers TMD PDF extracted from SIDIS data \cite{compass:2018weighted} was used to calculate the DY $q_T$ weighted asymmetry. Possible $Q^2$ evolution effects due to somewhat different hard scales of two measurements are neglected. The procedure is exactly the same followed for the 2015 data set, described in \cite{matousek:2018}. 
It makes use of the unpolarized PDFs from CTEQ~5D global fit\,\cite{cteq5,lhapdf6} and the FFs from DSS~07 global fit\,\cite{dss:2007}, to determine the first transverse moments of u- and d-quark Sivers functions $\fSiv{(1)\mr{u/d}}(x)$. 
A parametrization $x \fSiv{(1)q}(x) = a_q \, x^{b_q} \, (1-x)^{c_q}$ is used to fit the weighted asymmetries in SIDIS for both $h^+$ and $h^-$. 
The projection to the DY case is obtained using \refeq{eq:siv} and assuming no contribution of the sea quarks in the flavour sums, and the sign-change of the Sivers function between SIDIS and DY~\cite{collins:2002}: 
\beq
	\label{eq:sivapp}
	\A{T}{\sin\phiS\frac{\qTscal}{M_\mr{p}}}(x_\pi,x_N) 
		\approx -2 \frac{\fSivb{(1)\mr{u}}(x_N)}{f_{1,\mr{p}}^\mr{u}(x_N)}
		= 2 \frac{\fSivb{(1)\mr{u}}(x = x_N)|_\mr{SIDIS}}{f_{1,\mr{p}}^\mr{u}(x_N)}.
\eeq
Looking at the $x_N$ vs $x_\pi$ distribution in Fig.~\ref{fig:xbxN}, one can see that COMPASS covers the valence region of both p and $\pi^-$; therefore, this assumption is fairly justified. An additional suppression comes from the fractional quark charge. The advantage of the assumptions above is the cancellation in the ratio of the pion PDF, that is still poorly known. The unpolarized PDF used, $f_{1,\mr{p}}^\mr{u}$, is the same adopted in the SIDIS case. The result is compared to the weighted Sivers asymmetry extracted from the combined 2015 + 2018 DY sample in Fig~\ref{fig:ResSivPredDY}. 

The systematic uncertainty is discussed in detail in \cite{matousek:2018}.
The significance of the result does not yet allow one to draw conclusions on the sign-change hypothesis. Currently, the second part of 2018 data set is being analyzed to reduce the statistical error. 
\subsection{Boer-Mulders function}
Using the weighted TSAs one can also access other TMDs, like the pion Boer--Mulders (BM) TMD PDF. This function was never extracted before. It can be independently obtained from COMPASS results using the $\A{T}{\sin(2\phi-\phiS) \frac{\qTscal}{M_\pi}}$ asymmetry shown in Fig.~\ref{fig:wasymcom} using \refeq{eq:transv}: 
\beq
    \begin{split}
	\A{T}{\sin(2\phi-\phi_S) \frac{\qTscal}{M_\pi}}(x_\pi,x_N)
		&= - 2 \frac{ \sum_q e_q^2 \bigl[ 
						h_{1,\pi}^{\perp (1) \bar{q}}(x_\pi) \, 
						h_{1,\mr{p}}^q (x_N) 
					+ \qqswap \bigr]}
			 {\sum_q e_q^2 
				\bigl[ f_{1,\pi}^{\bar{q}}(x_\pi) \, f_{1,\mr{p}}^q(x_N) + \qqswap \bigr]}
		\\
		&\approx	- 2  \frac{ e_\mr{u}^2 \, h_{1,\pi}^{\perp (1) \bar{\mr{u}}}(x_\pi) \, 
							h_{1,\mr{p}}^\mr{u} (x_N) }
			 {\sum_{q=\mr{u,d,s}} e_q^2 
				\bigl[ f_{1,\pi}^{\bar{q}}(x_\pi) \, 
				    f_{1,\mr{p}}^q(x_N) + \qqswap \bigr]}.
    \end{split}
\eeq
where the Boer--Mulders and Transversity PDFs of sea quarks are assumed to be zero and only u, d and s quark contributions are considered in the denominator. Under these assumptions, all the remaining quantities on the left hand side of the equation can be found in the literature, except for the first transverse moment of the Boer--Mulders function. 

We make use of CTEQ~5D proton PDFs and the GRV-PI pion PDF\cite{grv:1992} from the LHAPDF library \cite{lhapdf6} at $Q^2 = 25\,(\mr{GeV}/c)^2$, which is compatible with the scale of COMPASS DY measurements in the HM range. The valence pion PDF depends on $x_\pi$, while the proton PDFs are nearly constant in the explored range in $x_N$ (shown in Fig~\ref{fig:xbxN}).
For the Transversity TMD PDF, a point-by-point extraction from COMPASS data~\cite{Martin:2014wua} was used. 
In this work the hard scale dependence of the Transversity TMD PDF has been neglected.
In the extraction of $h_{1,\pi}^{\perp (1) \bar{\mr{u}}}$ the uncertainties on the unpolarized proton and pion PDFs are neglected. The final errors account only for the statistical uncertainties of the Transversity PDF and the uncertainty of the weighted asymmetry. 

The first transverse moment of valence Boer--Mulders function of pion is shown in Fig.~\ref{fig:ResBMDY}. The plot includes two sets of error bars, statistical only and combined statistics and systematics errors quadratic sum. The results are compared with a parametrization from Ref.~\cite{boer:1999}.

\end{document}